\documentstyle[11pt,amssymb,epsf]{article}

\makeatletter
\@addtoreset{equation}{section}
\makeatother

\def\eq#1{(\ref{#1})}

\def\ben{\begin{equation}}
\def\een{\end{equation}}

\let\a=\alpha \let\b=\beta   \let\e=\epsilon
    
 \let\m=\mu \let\n=\nu

\let\C=\Chi 
\let\la=\label  
  
\def\nn{\nonumber} \def\bd{\begin{document}} \def\ed{\end{document}}
\def\ds{\documentstyle} \let\fr=\frac \let\bl=\bigl \let\br=\bigr
\let\Br=\Bigr \let\Bl=\Bigl
\let\bm=\bibitem
\let\na=\nabla
\let\pa=\partial \let\ov=\overline
\newcommand{\be}{\begin{equation}}
\newcommand{\ee}{\end{equation}}
\def\ba{\begin{array}}
\def\ea{\end{array}}
\def\ft#1#2{{\textstyle{{\scriptstyle #1}\over {\scriptstyle #2}}}}
\def\fft#1#2{{#1 \over #2}}
\def\del{\partial}
\def\vp{\varphi}
\def\sst#1{{\scriptscriptstyle #1}}
\def\oneone{\rlap 1\mkern4mu{\rm l}}
\def\td{\tilde}
\def\wtd{\widetilde}
\def\ie{\rm i.e.\ }
\def\dalemb#1#2{{\vbox{\hrule height .#2pt
        \hbox{\vrule width.#2pt height#1pt \kern#1pt
                \vrule width.#2pt}
        \hrule height.#2pt}}}
\def\square{\mathord{\dalemb{6.8}{7}\hbox{\hskip1pt}}}
\newcommand{\ho}[1]{$\, ^{#1}$}
\newcommand{\hoch}[1]{$\, ^{#1}$}
\newcommand{\bea}{\begin{eqnarray}}
\newcommand{\eea}{\end{eqnarray}}
\newcommand{\ra}{\rightarrow}
\newcommand{\lra}{\longrightarrow}
\newcommand{\Lra}{\Leftrightarrow}
\newcommand{\ap}{\alpha^\prime}
\newcommand{\bp}{\tilde \beta^\prime}
\newcommand{\tr}{{\rm tr} }
\newcommand{\Tr}{{\rm Tr} }
\def\0{{\sst{(0)}}}
\def\1{{\sst{(1)}}}
\def\2{{\sst{(2)}}}
\def\3{{\sst{(3)}}}
\def\4{{\sst{(4)}}}
\def\5{{\sst{(5)}}}
\def\6{{\sst{(6)}}}
\def\7{{\sst{(7)}}}
\def\8{{\sst{(8)}}}
\def\n{{\sst{(n)}}}
\def\cA{{{\cal A}}}
\def\cF{{{\cal F}}}
\def\tV{\widetilde V}
\def\tW{\widetilde W}
\def\tH{\widetilde H}
\def\tE{\widetilde E}
\def\tF{\widetilde F}
\def\tA{\widetilde A}
\def\im{{{\rm i}}}
\def\tY{{{\wtd Y}}}
\def\ep{{\epsilon}}
\def\vep{{\varepsilon}}
\def\R{\rlap{\rm I}\mkern3mu{\rm R}}
\def\bD{{{\bar D}}}

\def\R{\rlap{\rm I}\mkern3mu{\rm R}}
\def\bD{{{\bar D}}}
\def\R{{{\Bbb R}}}
\def\C{{{\Bbb C}}}
\def\H{{{\Bbb H}}}
\def\CP{{{\Bbb C}{\Bbb P}}}
\def\RP{{{\Bbb R}{\Bbb P}}}
\def\Z{{{\Bbb Z}}}
\def\bA{{{\Bbb A}}}
\def\bB{{{\Bbb B}}}
\def\bC{{{\Bbb C}}}
\def\bR{{{\Bbb R}}}
\def\bD{{{\Bbb D}}}
\def\bE{{{\Bbb E}}}
\def\bZ{{{\Bbb Z}}}
\def\Re{{{\frak{Re}}}}
\def\Im{{{\frak{Im}}}}
\def\cosec{{\,\hbox{cosec}\,}}
\def\Gm{{\Gamma_{\!\! -}}}
\def\Gp{{\Gamma_{\!\! +}}}
\def\stan{{standard }}
\def\nonstan{{supernumerary }}

\def\cosech{{\hbox{cosech}}}

\def\etcyc{{\hbox{and cyclic}}}
\def\btheta{{\bar\theta}}
\def\m{\mu}
\def\n{\nu}

\newcommand{\tamphys}{\it Center for Theoretical Physics,
Texas A\&M University, College Station, TX 77843, USA}
\newcommand{\umich}{\it Michigan Center for Theoretical Physics,
University of Michigan\\ Ann Arbor, MI 48109, USA}
\newcommand{\upenn}{\it Department of Physics and Astronomy,\\
University of Pennsylvania, Philadelphia,  PA 19104, USA}
\newcommand{\SISSA}{\it  SISSA-ISAS and INFN, Sezione di Trieste\\
Via Beirut 2-4, I-34013, Trieste, Italy}

\newcommand{\mitchell}{\it George P. and Cynthia W.
Mitchell Institute for Fundamental Physics,\\
Texas A\&M University, College Station, TX 77843-4242, USA}

\newcommand{\newton}{\it Isaac Newton Institute for Mathematical Sciences,\\
20 Clarkson Road,  University of Cambridge,
Cambridge CB3 0EH, UK}

\newcommand{\ihp}{\it Institut Henri Poincar\'e\\
  11 rue Pierre et Marie Curie, F 75231 Paris Cedex 05}

\newcommand{\damtp}{\it DAMTP, Centre for Mathematical Sciences,
 Cambridge University\\  Wilberforce Road, Cambridge CB3 OWA, UK}
\newcommand{\itp}{\it Institute for Theoretical Physics, University of
California\\ Santa Barbara, CA 93106, USA}

\newcommand{\auth}{
H. L\"u,\hoch{*} C.N. Pope\hoch{*} and E. Sezgin\hoch{\dagger}}

\def\ho{{\widehat\omega}}
\def\hh{{\widehat H}}
\def\hhp{{\widehat H'}}
\def\pb{{\bar\psi}}
\def\hp{{\widehat\psi}}
\def\eb{{\bar\epsilon}}

\begin{document}
\begin{flushright}
\hfill{MIFP-03-10}\\
\hfill{ \bf hep-th/0305242}
\end{flushright}

\vspace{15pt}

\begin{center}

{\large {\bf Yang-Mills-Chern-Simons Supergravity}}

\vspace{15pt}

\auth

\vspace{7pt}
{\hoch{\ddagger}\mitchell}

\vspace{75pt}

\underline{ABSTRACT}
\end{center}

\vspace{15pt}

      $N=(1,0)$ supergravity in six dimensions admits AdS$_3\times
S^3$ as a vacuum solution. We extend our recent results presented
in hep-th/0212323, by obtaining the complete $N=4$
Yang-Mills-Chern-Simons supergravity in $D=3$, up to quartic
fermion terms, by $S^3$ group manifold reduction of the six
dimensional theory. The $SU(2)$ gauge fields have Yang-Mills
kinetic terms as well as topological Chern-Simons mass terms.
There is in addition a triplet of matter vectors. After
diagonalisation, these fields describe two triplets of
topologically-massive vector fields of opposite helicities.  The
model also contains six scalars, described by a $GL(3,R)/SO(3)$
sigma model. It provides the first example of a three-dimensional
gauged supergravity that can obtained by a consistent reduction of
string-theory or M-theory and that admits AdS$_3$ as a vacuum
solution.  There are unusual features in the reduction from
six-dimensional supergravity, owing to the self-duality condition
on the 3-form field.  The structure of the full equations of
motion in $N=(1,0)$ supergravity in $D=6$ is also elucidated, and
the role of the self-dual field strength as torsion is exhibited.

{\vfill\leftline{}\vfill
\vskip 10pt \footnoterule
{\footnotesize \hoch{*}
Research supported in part by DOE grant DE-FG03-95ER40917
\vskip -12pt} \vskip 14pt
{\footnotesize \hoch{\dagger}
Research supported in part by NSF Grant PHY-0070964
\vskip -12pt}  \vskip  14pt
}

\pagebreak
\setcounter{page}{1}

\tableofcontents
\addtocontents{toc}{\protect\setcounter{tocdepth}{2}}


\section{Introduction}


      A model of considerable interest in the context of
AdS$_3$/CFT$_2$ correspondence is the AdS$_3\times S^3$
compactification of the $N=(2,0)$ supergravity in $D=6$ coupled to 21
tensor multiplets \cite{malda,ms}. This model arises in Type IIB
string on $K3$ and it has $16$ real supersymmetries \cite{pkt6}. The
AdS$_3\times S^3$ compactification preserves all supersymmetries, and
it was determined in \cite{dkss} that the propagating massless
Kaluza-Klein spectrum consists of 21 hyper-multiplets and a special
vector multiplet comprising $SO(4)$ Yang-Mills fields, 6 additional
vector fields and 26 scalars.  As a first step in finding the AdS$_3$
supergravity with 16 supersymmetries describing the couplings of this
system, we recently studied \cite{lps} the simpler problem of
obtaining a similar AdS$_3$ supergravity with $8$ supersymmetries by
means of an $S^3$ group manifold reduction \cite{scsc}
\footnote{By a group manifold reduction, we mean a dimensional
reduction on a group manifold in which only those fields associated
with left-invariant harmonics on the group are retained.  Provided that
one keeps all such fields, the consistency of the reduction is
guaranteed.}
 of $(1,0)$ supergravity \cite{ns,r1,ns1,bss,ns97,ric1}.  In
particular, in \cite{lps} we obtained the bosonic Lagrangian and
the supersymmetry transformations of the fermions in three
dimensions. In this paper, we shall extend these results to obtain
the fermionic part of the resulting supergravity Lagrangian and
the supersymmetry transformations of the bosons as well. There are
subtleties in this process due to the presence of a self-duality
equation in the pure $(1,0)$ theory in six dimensions. As we shall
see, the use of a $6D$ Lagrangian for the fermions will require an
ansatz for the self-dual field strength that contains special
fermionic bilinears terms. While the idea of using the fermionic
part of Lagrangian in reduction schemes has been considered
before, these schemes apparently have not been carried out fully
to find supersymmetric results in lower dimensions. The procedure
we have found in this paper fills this gap and it can have
applications for the reductions of other chiral supergravity
theories with self-dual field strengths.

       Some of the salient features of the $3D$ supergravity
we have obtained are:
\begin{itemize}

\item[(1)] The theory admits AdS$_3$ as vacuum (this
is the first instance of a group manifold type reduction giving rise
to such a feature.)

\item[(2)] The theory has two sets of vector fields; a triplet
coming from the metric, which we refer to as A-type vector fields,
a triplet coming from the self-dual 3-form, which we refer to as
B-type vector fields. The A-type vector fields are $SU(2)$ gauge
fields with Yang-Mills kinetic term and a topological mass term
(hence the terminology of Yang-Mills-Chern-Simons supergravity),
while the B-type vector fields are $SU(2)$ matter fields which mix
with the $SU(2)$ gauge fields. After diagonalisation, these fields
describe two triplets of topologically-massive vector fields of
opposite helicities.

\item[(3)] Although there exist many gauged supergravities in $D=3$
that admit an AdS$_3$ vacuum, our theory is the only one known so far
that has a consistent string or M-theory origin.\footnote{Recently,
a paper has appeared which initiates the direct construction in three
dimensions of a broad class of gauged supergravities \cite{nsr}.  This
construction, which to date presents the bosonic action only, presumably
should encompass our example that arises from dimensional reduction.}

\end{itemize}

      In this paper, we also elucidate the structure of the pure
$(1,0)$ supergravity theory by providing its full field equations
in various forms, and we highlight the occurrence of the self-dual
3-form field strength as bosonic torsion in these equations. This
fact was pointed out long ago in \cite{bss} up to quartic fermion
terms in the action. We extend these results here by the inclusion
of all the quartic fermion terms as well.

     The first part of this paper, contained in Section 2, deals with
the structure of the pure $(1,0)$ supergravity in $D=6$. In the
second part, contained in Section 3, we describe the $S^3$ group
manifold reduction of the theory, including its supersymmetry
transformation rules, and we also determine the representation
content of the vector fields, showing that they describe two
triplets of spin-1 states with opposite helicities. In the
concluding Section 4, we comment further on our results.


\section{The Complete $(1,0)$ Supergravity in $D=6$}


The pure $(1,0)$ supergravity in $D=6$ was considered in \cite{r1}
at the level of the purely bosonic field equations, and the lowest
order in fermions gravitino field equation and supersymmetry
transformation rules. The complete field equations can be obtained
by suitable truncations of the matter coupled versions given in
\cite{ns1,bss}, or by imposing self-duality condition on the
3-form field strength {\it after} the variation of an action
(which is not supersymmetric but it does yield supersymmetric
field equations by this procedure) \cite{ric1}.

In this section, we begin with the description of the complete field
equations with an emphasis on how the the self-dual 3-form field
strength arises as torsion. This property, noted at lowest order in
fermionic contributions in \cite{bss}, will be shown to hold at the
level of the complete field equations. We then proceed to truncate the
equations to lowest order in fermionic terms consistent with
supersymmetry, which requires that the fermionic bilinear
contributions to the bosonic field equations be kept. We use these
results in Section 3 to perform the $S^3$ group manifold reduction.


\subsection{The Complete  Field Equations and Bosonic Torsion}


The $D=6$, $(1,0)$ supergravity multiplet consists of the vielbein,
2-form potential with self-dual field strength and a gravitino which
is symplectic Majorana-Weyl spinor in doublet representation of the
R-symmetry group $Sp(1)$. As is well known, a manifestly covariant
action containing these fields alone cannot be written down due to the
the self-duality condition. However, the coupling of this multiplet to
a tensor multiplet, consisting of a two-form potential with anti-self
dual field strength, a dilaton and anti-chiral symplectic-Majorana
spinor, does admit a Lagrangian formulation. Indeed, the complete
Lagrangian, field equations and supersymmetry transformation rules for
the coupled system have been constructed in \cite{ns1}. Starting from
these field equations and transformation laws, we can obtain the
corresponding ones for the pure supergravity theory by setting the
dilaton and the tensor-multiplet spinor to zero, and imposing
self-duality condition on the supercovariant 3-form field
strength. The resulting supersymmetry transformation rules are
\bea
\delta { e}_\mu{}^a &=& {\bar\epsilon}\,
\Gamma^a{\psi}_\mu\ ,\la{s1}\\[5pt]
\delta B_{\mu\nu}&=& -{\bar \epsilon}
\Gamma_{[\mu}{\psi}_{\nu]}\ ,\la{s2}\\[5pt]
\delta {\psi}_\mu &=& {\cal D}_\mu(\omega_+)\,\epsilon\ ,
\la{s3}
\eea
where $\Gamma_7\psi=\psi$, and
\bea {\cal D}_\mu (\omega_+)\epsilon &=& \left(\partial_\mu  +
\ft14
\omega^+_{\mu ab} \Gamma^{ab}\right) \epsilon\ , \la{opm}\\[5pt]
\omega^{\pm}_{\mu ab}&=&\ho_{\mu ab}\pm \hh^+_{\mu ab}\ .
\eea
Here the supercovariant spin connection and 3-form field strengths are
defined as
\bea
\ho_\m{}^{ab} &=& \omega_\mu{}^{ab}(e)+\kappa_\mu{}^{ab}\
,\\[5pt]
\hh_{\mu\nu\rho} &=& H_{\mu\nu\rho}+\ft32
{\bar\psi}_{[\mu}\Gamma_\nu\psi_{\rho]}\ ,
\eea
where $\omega_\mu{}^{ab}(e)$ is defined in the Appendix B and
\bea \kappa_{\mu ab} &=& \pb_\mu\Gamma_{[a}\psi_{b]}
+\ft12\pb_a\Gamma_\mu\psi_b\ , \\[5pt]
H_{\mu\nu\rho}&=& 3\partial_{[\mu} B_{\nu\rho]}\ .
\eea
The (anti)self-dual projections are defined as $ H_{abc}^{\pm} =
\ft12(H_{abc} \pm \ft1{3!}\,\epsilon_{abcdef} H^{def})$. Further
notations and conventions can be found in Appendix A.

The field equations can be obtained either from the closure of the
supersymmetry transformations \eq{s1}-\eq{s3}, or directly from the
field equations given in \cite{ns1} of the $(1,0)$ supergravity
coupled to a single tensor multiplet by means of a consistent
truncation triggered by setting the dilaton and the tensor-multiplet
spinor to zero, and imposing self-duality condition on
$\hh_{\mu\nu\rho}$.  We find that the resulting full field equations
of of the pure $(1,0)$ supergravity in six dimensions can be expressed
as follows:

\bea
\hh^-_{\mu\nu\rho} &=& 0\ ,\la{e1}\\[5pt]
\Gamma^\mu \hp_{\mu\nu}&=& 0\ ,\la{e2}\\[5pt]
{\widehat R}_{\mu\nu} &=&0\ , \la{e3}\eea

where the supercovariant gravitino curvature is defined as

\be \hp_{\mu\nu} \equiv{\cal D}_\mu(\omega_+)\psi_\nu -{\cal
D}_\nu(\omega_+)\psi_\mu\ ,\ee

and supercovariant generalized Ricci tensor is given by

\be {\widehat R}_{\mu\nu} \equiv  R_{\mu\nu}(\omega_+)
-\pb^a\Gamma_a \hp_{\mu\nu} -\pb^a\Gamma_\mu \hp_{\nu a} +2 \pb^a
\Gamma^b \psi_\nu\hh^+_{\mu ab}\ . \ee

The definition and properties of the Ricci tensor
$R_{\mu\nu}(\omega_+)$ are given in Appendix B. Note that
neither ${\widehat R}_{\mu\nu}$ nor $R_{\mu\nu}(\omega_+)$ are
symmetric. Taking the symmetric and antisymmetric part of \eq{e3},
we find
\bea R_{(\mu\nu)}(\omega_+) -\pb^a\Gamma_{(\mu}\hp_{\nu)a} +2
\pb^a\Gamma^b \psi_{(\mu}\hh^+_{\nu)ab} &=&0\ ,\la{s}\\[5pt]
R_{[\mu\nu]}(\omega_+)-\pb^a\Gamma_a\hp_{\mu\nu}
-\pb^a\Gamma_{[\mu}\hp_{\nu]a} -2 \pb^a\Gamma^b
\psi_{[\mu}\hh^+_{\nu]ab} &=&0\ .\la{as}\eea
Computing the antisymmetric part of $R_{\mu\nu}(\Gamma)$ from its
definition yields the result \eq{ricci} given in Appendix B.
Using this result in \eq{as} we find
\be{\cal D}_\lambda(\Gamma)\hh_{\mu\nu}^{+\ \lambda} -\ft12
\pb_a\Gamma^a\hp_{\mu\nu} -\pb^a\Gamma_{[\mu}\hp_{\nu]a}=0\ .
\la{e33}\ee
It is gratifying to check that not only this equation is
supercovariant but it also follows from the curl of the self-duality
equation \eq{e1}, namely from
$\epsilon^{\mu\nu\rho\sigma\lambda\tau}{\cal
D}_\mu{(\Gamma)}\hh^-_{\nu\rho\sigma}=0$. Therefore, in a formalism
where we work with $\hh^+_{\mu\nu\rho}$, we can take \eq{e33} to be
the field equation for the 2-form potential. Note also that making use
of the formulae \eq{rr2} and \eq{gr2} given in Appendix B, the
supercovariant Einstein equation \eq{s} can be written as
\be R_{(\mu\nu)}(\ho) = \hh^+_{\mu ab}\hh_\nu^{+\
ab}+\pb^a\Gamma_{(\mu}\hp_{\nu)a} -2 \pb^a\Gamma^b
\psi_{(\mu}\hh^+_{\nu)ab} \ .\la{ss} \ee
In checking the super-covariance of various equations we have
encountered above, it is useful to note the following results:
\bea \delta\ho_{\mu ab} &=& \eb\Gamma_{[a}\hp_{b]\mu}-\ft12
\eb\Gamma_\mu\hp_{ab}+\eb\Gamma^c\psi_\mu \hh^+_{abc}\
,\\[5pt]
\delta \hh^+_{\m ab} &=& \ft32 \eb \Gamma_{[\mu} \hp_{ab]}
+\eb\Gamma^c\psi_\mu \hh^+_{abc}\ . \la{dh1}\eea
In deriving  the second equation, we have used the self-duality
equation \eq{e1} and the Fierz identity
\be \Gamma_a\psi_{[\mu}\pb_{\nu}\Gamma^a\psi_{\rho]}=0\ . \ee
From \eq{dh1} and the gravitino field equation \eq{e2}, we also
find
\be \delta \hh^+_{abc} = \ft32 \eb \Gamma_{[a} \hp_{bc]}\
.\la{dh2}\ee
The self duality of the right hand sides is ensured by the
gravitino field equation \eq{e2}.

To summarize, the full field equations of the pure $N=(1,0)$
supergravity are given by \eq{e1}, \eq{e2} and \eq{e3}, in which
the self-dual and supercovariant 3-form field strength manifestly
arises as torsion, or equivalently by \eq{e2}, \eq{e33} and
\eq{ss}.


\subsection{Truncation of the Quartic Fermions }


The bosonic field equations given in the previous section, \eq{e2},
\eq{e33} and \eq{ss}, have quadratic and quartic in fermion
contributions. To simplify the $SU(2)$ reduction which will be
performed in the next section, the quartic fermion terms in these
equations and cubic fermion terms in the supersymmetry transformation
rules (counting the supersymmetry parameter as one of the fermionic
fields) can be truncated such that the field equations are
supersymmetric up to quartic fermion terms. While the self-duality
equation \eq{e1} must be implemented fully in this process, the
equivalent second order field equation \eq{e33} can be consistently
truncated to keep terms up to quadratic order in fermions.

Implementing the truncation procedure outlined above to the field
equations \eq{e2}, \eq{e33} and \eq{ss}, we find:
\bea \Gamma^\mu \psi_{\mu\nu}&=&\ft14
\Gamma^\mu\Gamma^{ab}\psi_\mu H^+_{\nu ab}\ ,\la{ne1}\\[5pt]
\nabla_\rho H^{\mu\nu\rho}_+ &=& \ft18 \nabla_\rho \Big(
\pb_\lambda \Gamma^{\lambda\tau\mu\nu\rho}\psi_\tau -3! \pb^{[\mu}
\Gamma^\nu\psi^{\rho]}\Big) +\ft32 \pb^a
\Gamma_{[\mu}\psi_{\nu a]} \nn\\[5pt]
&& -\pb^a\Gamma_{[\mu}\psi^b H^+_{\nu] ab} -\pb^a\Gamma^b
\psi_{[\mu} H^+_{\nu] ab}-\pb_c\Gamma^c \psi^a H^+_{\mu\nu a}\
,\la{ne2}\\[5pt]
R_{\mu\nu}&=& H_{\mu ab}^+ H_\nu^{+\ ab} -\nabla_{(\mu}
\left(\pb_{\nu)} \Gamma^a\psi_a\right)-\nabla_{\rho}
\left(\pb^\rho \Gamma_{(\mu}\psi_{\nu)}\right)\nn\\[5pt]
&&+\pb^a\Gamma_{(\mu} \psi_{\nu)a} +\ft14 \pb_c
\Gamma^{ab}{}_{(\mu} \psi^c H^+_{\nu)ab} -\ft14 \pb^a
\Gamma^{bc}{}_{(\mu} \psi_{\nu)} H^+_{abc}\nn\\[5pt]
&&+\ft32 \pb^a \Gamma^b \psi_{(\mu} H^+_{\nu)ab}- \pb^a
\Gamma_{(\mu} \psi^b H^+_{\nu)ab} -\ft12 g_{\mu\nu} \pb^a \Gamma^b
\psi^c H^+_{abc}\ ,\la{ne3} \eea
where $R_{\mu\nu}\equiv R_{\mu\nu}(\omega(e))$, the covariant
derivative $\nabla_\mu$ is torsion free and
\be \psi_{\mu\nu}=\left(
\partial_\mu+\ft14\omega_\mu{}^{ab}(e)\Gamma_{ab}\right)\psi_\nu
-\mu\leftrightarrow \nu\ .\ee
The supersymmetry transformation up to cubic fermions, on the
other hand, take the form
\bea \delta { e}_\mu{}^a &=& {\bar\epsilon}\,
\Gamma^a{\psi}_\mu\ ,\la{s1n}\\[5pt]
\delta B_{\mu\nu}&=& -{\bar \epsilon}
\Gamma_{[\mu}{\psi}_{\nu]}\ ,\la{s2n}\\[5pt]
\delta {\psi}_\mu &=& \Big[\partial_\mu +\ft14
\Big(\omega_\mu{}^{ab}(e)+H_\mu^{+\
ab}\Big)\Gamma_{ab}\Big]\,\epsilon\ , \la{s3n} \eea
The above field equations are rather complicated. The strategy now
is to determine if they can be derived from the variation of an
action by a consistent procedure. A natural candidate is to start
from the Lagrangian \cite{ric1}
\bea e^{-1}{\cal L}= \ft14 R -\ft1{12}
H_{\mu\nu\rho}H^{\mu\nu\rho}
-\ft14\pb_\mu\Gamma^{\mu\nu\rho}\psi_{\nu\rho}-
\ft1{24}\psi^\lambda\Gamma_{[\lambda}
\Gamma^{\mu\nu\rho}\Gamma_{\tau]}\psi^\tau H_{\mu\nu\rho}\ ,
\la{6d}\eea
which is the up to quartic fermion Lagrangian of $(1,0)$
supergravity coupled to a tensor multiplet and where the dilaton
and spinor fields of that multiplet are set to zero. Varying with
respect to the graviton, gravitino and the 2-form potential, and
{\it after} the variation making use of the full supersymmetric
self-duality condition \eq{e1} which can be expressed as
\be
H_{\mu\nu\rho}=H^+_{\mu\nu\rho}
-\ft18\psi_\lambda\Gamma^{[\lambda}\Gamma_{\mu\nu\rho}
\Gamma^{\tau]}\psi_\tau\, \la{sd2}
\ee
we find the field equations
\bea \Gamma^{\mu\nu\rho}\psi_{\nu\rho}&=&
-2H_+^{\mu\nu\rho}\gamma_\nu\psi_\rho\ ,\la{f1}\\[5pt]
\nabla_\mu H_+^{\mu\nu\rho}&=& -\ft18 \nabla_\mu\Big(
\pb^\lambda\Gamma_{[\lambda}\Gamma^{\mu\nu\rho}
\Gamma_{\tau]}\psi^\tau\Big)\,\la{f2}\\[5pt]
R_{\mu\nu} &=& H_{\mu ab}^+ H_\nu^{+\ ab} -\nabla_{(\mu}
\left(\pb_{\nu)} \Gamma^a\psi_a\right)-\nabla_{\rho}
\left(\pb^\rho \Gamma_{(\mu}\psi_{\nu)}\right)\nn\\[5pt]
&&-\pb_c\Gamma^{ca}{}_{(\mu}\psi_{\nu)a} +\ft12\pb_{(\mu}
\Gamma_{\nu)}{}^{ab} \psi_{ab}+24\pb^a\Gamma^b\psi_{(\mu}
H^+_{\nu)ab} \ .\la{f3}\eea
In particular, the coefficient $-1/8$ in \eq{f2} results from two
contributions of the same type; one from the variation of the Pauli
coupling term in the Lagrangian and another from the substitution
\eq{sd2}.

The question now is whether these equations agree with \eq{ne1},
\eq{ne2} and \eq{ne3} which were derived from first principles. A
lengthy computation in which the field equations are used repeatedly
shows that this is indeed the case.

Note that the self-dual part of $H$ is automatically picked up in the
Pauli coupling in the Lagrangian. Yet, the above Lagrangian is not
supersymmetric because the kinetic term contains the anti-selfdual
part as well, and indeed the self-duality equation does not follow
from this Lagrangian. Nonetheless, we have proven above that the field
equations following from this Lagrangian {\it followed} by the use of
the self-duality equation \eq{sd2} gives the correct equation of
motion derived from first principles. The above result will simplify
considerably the $SU(2)$ reduction described in the next section.


\section{The $S^3$ Group Manifold Reduction }


We begin by explaining our strategy for handling the self-duality
condition in performing the $S^3$ group manifold reduction. The
resulting 3D supergravity Lagrangian is given by \eq{pa},
\eq{bosonic} and \eq{fermionic}, and the supersymmetry
transformations in \eq{3}, \eq{4}, \eq{7}, \eq{8}, 
\eq{10}, \eq{chi} and \eq{deltaB}.


\subsection{The Strategy for the Reduction of the Self-Duality Condition}


The most straightforward, though certainly not the most economical,
way to perform Scherk-Schwarz reduction of the model described above
on $S^3$ group manifold down to $D=3$ is to first reduce the field
equations, and then find a $D=3$ Lagrangian from which they can be
derived. In the case of bosonic field equations, this was done in
\cite{lps}. To obtain the supergravity up to quartic fermions,
however, a considerably simpler simpler way to proceed is to make use
of the Lagrangian \eq{6d}. To do so, we first observe that:

1) The gravitino field equation obtained from the Lagrangian \eq{6d}
is evidently the same regardless of whether the self-duality equation
\eq{sd2} is used before or after the variation of the Lagrangian. This
is due to the fact that the anti-selfdual part of the 3-form field
strength is automatically dropped out of the Pauli coupling, while the
H-kinetic term, of course, does not effect the gravitino field
equation.

2) The Einstein equation obtained from the variation of the Lagrangian
\eq{6d} is also the same regardless of whether \eq{sd2} is used before
or after the variation. This is not obvious but we have checked that
the substitution \eq{sd2} into the Einstein's equation does not yield
fermionic contributions over and above those which arise from the last
term in \eq{6d}.

It follows from the above observations that a candidate for the
$S^3$ reduced action in $D=3$ is
\be S^{3D}= \int d^3x\, {\cal L}_B^{3D} +\int d^6 x\, {\cal L}_F\
, \la{pa}\ee

where ${\cal L}_B^{3D}$ is the 3Dbosonic Lagrangian obtained in
\cite{lps} (see \eq{bosonic}) and

\be {\cal L}_F = -\ft14\, {\widehat e}\, \hp_A\Gamma^{ABC}\hp_{BC}
-\ft12\,{\widehat e}\, \hp_A \Gamma_B\hp_C\, \hh_+^{ABC}\ ,\la{f}
\ee
where
\bea \hh_{ABC} &=& 3 {\widehat\partial}_{[A} {\widehat B}_{BC]}\ ,
\quad\quad {\widehat\partial}_A={\widehat e}_A{}^M
\partial_M\ ,\quad\quad\ \  A,M=0,1,...,5\ ,\nn\\[5pt]
\hp_{AB}&=&\Big[ \left( {\widehat\partial}_A+\ft14{\ho}_A{}
^{BC}(e)\Gamma_{BC}\right)\hp_B+\ho_{AB}{}^C(e)\,\hp_C-
A\leftrightarrow B\Big]\ , \la{32c} \eea
Note that in this section $(M,A)$ to denote the world and tangent
space indices in $6D$ and hats only refer to six dimensional
quantities and {\it not} supercovariantizations as they did in the
previous section. There should be no confusion in this notation
because we shall never use supercovariantized objects in this section
but rather explicitly write the terms required for
supercovarianizations if need be.

It will prove to be useful to express the Lagrangian \eq{f} as

\be {\cal L}_F = -\ft14\, {\widehat e}\, \hp_A\Gamma^{ABC}\hp_{BC}
-\ft1{24}\,{\widehat e}\,{\widehat X}^{ABC}\, \hh^+_{ABC}\
,\la{ff} \ee
where
\be {\widehat X}^{ABC}\equiv  \hp^D \Gamma_{[D}
\Gamma^{ABC}\Gamma_{E]} \hp^E \equiv e^{-\phi} X^{ABC}\ .
\la{x}\ee
Note that ${\widehat X}^{ABC}$ is ant-self-dual. In view of the
arguments given above, this Lagrangian yields the gravitino and
Einstein's field equations straightforwardly. The crucial check
remaining is to establish that reduction of the field equation for
the self-dual potential \eq{f2}, which in the notation of this
section we write as
\be{\widehat\nabla}_A \hh_+^{ABC} = -\ft18\,{\widehat\nabla}_A
{\widehat X}^{ABC}\ , \la{be}\ee
(here ${\widehat\nabla}$ is covariant derivative with respect to
the ordinary spin connection $\ho(e)$) agrees with that obtained
from our proposed action \eq{6d}. That check will ensure that the
ansatz for the self-dual field strength has the correct fermionic
bilinear terms. This we shall do in the remainder of this section.


\subsection{The Reduction Ansatz}


We begin with the introduction of  the left-invariant $SU(2)$
1-forms $\sigma^\a$, which satisfy the Maurer-Cartan algebra
\be
d \sigma^\a = -\ft12 f^\a{}_{\beta\gamma}\, \sigma^\beta\wedge
\sigma^\gamma\,,\label{cartan}
\ee
where $f^\a{}_{\beta\gamma}$ are the $SU(2)$ structure constants.
The Kaluza-Klein metric reduction ansatz will then be given by
\be d\hat s^2 = e^{2\a\, \phi}\, ds^2 + \fft{4}{g^2}\, e^{2\beta\,
\phi}\, h_{\a\beta}\, \nu^\a\, \nu^\beta\,,\label{metans} \ee
where $\phi$ is the ``breathing-mode'' scalar, $h_{\a\beta}$
denotes the remaining $n$-dimensional scalar fields (with the
symmetric tensor $h_{\a\beta}$ being unimodular), and $\nu_\a$ is
given by
\be \nu^\a \equiv \sigma^\a - g\,A^\a\ .\la{nu}\ee
The constants $\a$ and $b$ are chosen to be
\be \a=-3\beta=\sqrt {3\over 2}\ ,\ee
such that one obtains the standard Hilbert-Einstein term in $D=3$.
In \eq{nu}, $A^\a$ denotes the $SU(2)$ Yang-Mills potentials
corresponding to the right-acting $SU(2)$ isometry of the
3-sphere.

It  is convenient to work in a vielbein basis, which we take to be
\be \hat e^a = e^{\a\, \phi}\, e^a\,,\qquad \hat e^i = 2 g^{-1}\,
e^{\beta\, \phi}\,
 L^i_\a\, \nu^\a\ ,\qquad a,i=1,2,3\ ,\label{vielbein}
\ee
where $L^i_\a$ is the vielbein on $SL(3,R)/SO(3)$.
Here $e^a$ is a vielbein basis for the $3$-dimensional metric
$ds^2$, and $L^i_\a$ is a ``square root'' of $h_{\a\beta}$, and so
\be h_{\a\beta} = L^i_\a\, L^i_\beta\,,\qquad \det(L^i_\a)=1\,.
\ee
More explicitly,
\be
{\widehat e}_M{}^A = \left(%
\begin{array}{cc}
  e^{\alpha\phi} e_\mu{}^a & \ \ -\ft2{g}e^{\beta\phi}
A_\mu^\alpha L_\alpha^i \\
  0 & \ \ \ft2{g}e^{\beta\phi}L_\alpha^i \\
\end{array}%
\right)\ , \quad\quad {\widehat e}^M{}_A = \left(%
\begin{array}{cc}
  e^{-\alpha\phi} e^\mu{}_a & \ \ 0 \\
  e^{-\alpha\phi}e^\mu{}_a A_\mu^\alpha & \ \
\ft{g}2e^{-\beta\phi}L^\alpha_i \\
\end{array}%
\right) \ee
Note that ${\widehat e}\,_\alpha{}^a=0$ and ${\widehat
e}\,^\mu{}_i=0$. Defining the Yang-Mills field strengths $F^\a =
dA^\a + \ft12g\, f^\a{}_{\beta\gamma}\, A^\beta\wedge A^\gamma$,
we have:
\bea D\, F^\a &\equiv& dF^\a + g\,f^\a{}_{\beta\gamma}\,
A^\beta\wedge
F^\gamma=0\,,\nn\\
D\, \nu^\a &\equiv& d\nu^\a + g\,f^\a{}_{\beta\gamma}\,
A^\beta\wedge \nu^\gamma = -g\, F^\a -\ft12\,
f^\a{}_{\beta\gamma}\, \nu^\beta\wedge \nu^\gamma\,. \eea
It is also useful to define the Yang-Mills covariant exterior
derivative acting on the scalars $L^i_\a$:
\be D\, L^i_\a \equiv d L^i_\a - g\,f^\beta{}_{\gamma \a}\,
A^\gamma\,
 L^i_\beta\,.
\ee

The torsion-free spin connection $\hat \omega^A{}_B$, defined by
$d\hat e^A = -\hat\omega^A{}_B\wedge \hat e^B$ and
$\hat\omega_{AB}=-\hat\omega_{BA}$, where $A=1,...,6$ and the hats
denote six dimensional quantities, are found to be \cite{lps}
\bea \hat \omega_{cab} &=& e^{-\a\phi}\left[ \omega_{cab} + \a\,
\left(\del_b\phi\, \eta_{ac}\, - \del_a\phi\,
\eta_{bc}\right)\right]\ ,\nn\\
\hat\omega_{iab}&=&e^{-\ft73\phi} F_{ab i}\ ,\nn\\
\hat\omega_{abi}&=& -e^{-\ft73\phi} F_{ab i}\ ,\nn\\
\hat \omega_{ija} &=& e^{-\a\phi} \left( P_{a\, ij}-\ft13\a\,
\del_a\phi\, \delta_{ij}\right)\ ,\nn\\
\hat\omega_{aij} &=& e^{-\a\phi}\, Q_{a\, ij}\ ,\nn\\
\hat\omega_{kij}&=& \ft14\,g\,e^{\ft13\a\phi} \left(
C_{k,\,ij}-C_{i,\,jk}-C_{j,\,ki}\right)\ ,\la{sc}\eea
where we have defined
\bea L^{\a i}\, D_a L_\a^j &=& P_a^{ij}+Q_a^{ij}\ ,\quad\,
P_{ij}=P_{ji}\ ,\quad Q_{ij}=-Q_{ji}\ ,\\
T^{ij} &\equiv& L^i_\a\, L^j_\a\ ,\quad\quad\ \   C_{k,\,ij}
\equiv
T_k{}^\ell\, \epsilon_{\ell ij}\nn\\
F^i_{ab} &\equiv& L^i_\a\,F^\a\ ,\quad\quad\   L^\a_iL_\a^j=
\delta_i^j\ . \la{Tdef} \eea
It is also useful to define
\be {\cal D}_a\, L^i_\a \equiv D_a\,
L^i_\a  + Q_{a\, ij}\, L^j_\beta \,, \ee
from which it follows that
\be L^{\a i}\, {\cal D}_a \, L_\a^j = P_a^{ij}\ . \ee
Finally, we make following ansatz for the self-dual 3-form field
strength
\bea &&\hh^+_{abc} = m\, e^{\a\phi}\, \ep_{abc}(1+X) \,,\qquad
\hh^+_{abi} = -  e^{-\fft13 \a\phi}\, \ep_{ab}{}^c \,
(B_c^i+Y_c^i)\,,\nn\\[5pt]
&& \hh^+_{ijk} = m \, e^{\a\phi}\, \ep_{ijk}(1+X)\,,\qquad
\hh^+_{aij} = e^{-\fft13\a\phi}\, \ep_{ijk}\,
(B_a^k+Y_a^k)\,,\label{hcomp} \eea
where we have defined $B^i\equiv L^i_\a\, B^\a$, and $X,Y_a^i$ are
bilinear in fermions which are to be determined from the
requirement of the 3-form field equation following from the
proposed action \eq{pa} agrees with the $SU(2)$ reduction of the
field equation \eq{f2}. The ansatz above with $X$ and $Y_a^i$ set
to zero was used in \cite{lps} in obtaining the bosonic Lagrangian
in $D=3$. These fermionic modifications do not affect the
reduction of ${\cal L}_F$ in \eq{ff} but they are crucial in the
reduction of the field equation \eq{f2} and comparing the result
with that obtain from the variation of \eq{pa} with respect to the
field $B_a^i$. Finally, a word of caution with our notation: the
hats in this section, for example, in \eq{sc} and in \eq{hcomp} do
{\it not} refer to supercovariantization but merely to the six
dimensional natura of the quantities.


\subsection{The $3D$ Supergravity Lagrangian}


The total $3D$ supergravity Lagrangian is the sum \eq{pa}, where
the bosonic Lagrangian is given by \cite{lps}
\bea e^{-1}{\cal L}_B^{3D}&=& \ft14 R -\ft12
\partial_\mu\phi\partial^\mu\phi -\ft14
P_\mu^{ij}P^\mu_{ij}-\ft12
e^{-\ft83\a\phi}F_{\mu\nu}^iF^{\mu\nu}_i\nn\\[5pt]
&& -m^2 e^{4\a \phi}-\ft1{16} g^2 e^{\ft83\a\phi} \left(
T_{ij}T^{ij}-\ft12 T^2 \right)\nn\\[5pt]
&&-e^{\ft43 \a\phi} B_\mu^iB^\mu_i
-2g^{-1}\epsilon^{\mu\nu\rho}\Big( D_\mu B^\a_\nu  -2m
F_{\mu\nu}^\a\Big) B_\rho^\a \nn\\[5pt]
&& -8g^{-1}m^2 \epsilon^{\mu\nu\rho} \Big(A_\mu^\a\partial_\nu
A_\rho^\a+\ft13\epsilon_{\a\b\gamma}A_\mu^\a A_\nu^\b
A_\rho^\gamma\Big)\ .\la{bosonic}\eea
This is the result of reducing the bosonic Einstein equation and
the bosonic self-duality equation in $6D$ on $S^3$ group manifold
and constructing a Lagrangian which yields the resulting $3D$
bosonic field equations \cite{lps}. Our task is now to perform the
$SU(2)$ reduction of ${\cal L}_F$ given in \eq{ff}.

We begin by making an ansatz for the reduction of the gravitino
field. In doing so, we shall make use of the original treatment of
this problem in \cite{scsc}, and \cite{ss8} where it has been studied
further in the context of $S^3$ reduction of $D=11$ supergravity. One
technical aspect of the reduction is the diagonalization of the lower
dimensional gravitino and spinor kinetic terms. It is convenient to
treat the diagonalization problem after performing the $S^3$
reduction. Thus, we begin with the ansatz
\be \hp_a (x,y)= e^{-\ft12\a\phi(x)}\psi_a(x)\ , \quad\quad
\hp_i(x,y)= e^{-\ft12\a\phi(x)}\chi_i(x)\ , \la{a2} \ee
where the exponential factors are chosen such that the gravitino
kinetic term is canonical, i.e. with no dilaton prefactor.

Next, it is convenient to work out the $S^3$ reduction of the
gravitino curvature \eq{32c}. Using the vielbein basis
\eq{vielbein} and the spin connection \eq{sc}, we find:

\bea \hp_{ab} &=& e^{-\ft32\a\phi}\Big[\psi_{ab}
+\a\Gamma_a\Gamma^c\psi_b\,
\partial_c\phi +e^{-\ft43\a\phi}F_{ac}^i
\Big( \Gamma_i\Gamma^c\psi_b-2\delta^c_b\chi_i\Big)\Big]
\la{c1} \\[8pt]
\hp_{ij}&=& e^{-\ft32\a\phi}\Big[
\left(P_{cik}-\ft13\a\delta_{ik}\partial_c\phi\right)
\Gamma^k\Gamma^c\chi_j +\ft12 e^{-\ft43\a\phi}
F_{abi}\Gamma^{ab}\chi_j\nn\\
&&\quad\quad\quad\ \ +\ft18 g e^{\ft43\a\phi}\Big(
\left(C_{i,\,k\ell}-2C_{k,\,\ell i}\right)\Gamma^{k\ell}\chi_j
-4C_{k,\,ij}\chi^k\Big)\Big]
\la{c2}\\[8pt]
\hp_{ai} &=& e^{-\ft32\a\phi}\Bigg[{\cal D}_a\chi_i
+\ft12 P_{cij}\left(\Gamma^c\Gamma^j\psi_a +
2\delta_a^c\chi_j\right) \nn\\
&& \quad\quad\quad\ \ -\ft16\a \Big[\Gamma^c\Gamma_i\psi_a
+\left(5\delta^c_a-3\Gamma_a{}^c\right)
\chi_i\,\Big] \partial_c\phi \nn\\
&&\quad\quad\quad\ \ -\ft14 e^{-\ft43\a\phi}
F_{cd}^k\left(\delta_{ki}
\Gamma^{cd}\psi_a-2\Gamma^c\delta_a^d\Gamma_k\chi_i\right)\nn\\
&&\quad\quad\quad\ \ -\ft1{16} g e^{\ft43\a\phi}
\left(C_{i,jk}-2C_{j,\,ki}\right)\Gamma^{jk}\psi_a\Bigg]\ ,
\la{c3} \eea
where the antisymmetrizations in $[ab]$ and $[ij]$ on the right
hand sides are understood, and
\bea
{\cal D}_a\psi_b &=&\left(\partial_a
+\ft14\omega_{acd}\Gamma^{cd}+\ft14
Q_{ak\ell}\Gamma^{k\ell}\right)\psi_b +
\omega_{ab}{}^c\psi_c\ ,\la{d1}\\
{\cal D}_a\chi_i  &=&\left( \partial_a
+\ft14\omega_{acd}\Gamma^{cd}+\ft14 Q_{a
k\ell}\Gamma^{k\ell}\right)\chi_i +Q_{ai}{}^j\chi_j\ .\la{d2} \eea

Using the above results, we find the $S^3$ reduction of the
Lagrangian \eq{f} to be

\bea e^{-1}{\cal L}_F &=& -\ft12
{\bar\psi}_\mu\Gamma^{\mu\nu\rho}{\cal D}_\nu\psi_\rho
-{\bar\chi}_i\Gamma^i\Gamma^{\mu\nu}{\cal D}_\mu\psi_\nu
+\ft12 {\bar\chi}_i\Gamma^{ij}\Gamma^\mu{\cal D}_\mu\chi_j  \nn\\[5pt]
&& +\ft12( {\bar\psi}_\nu\Gamma^\mu\Gamma^\nu\Gamma^i\chi^j
+{\bar\chi}_k\Gamma^{ki}\Gamma^\mu\chi^j) P_{\,\mu ij} -\ft23
\alpha
\left({\bar\psi}_\nu\Gamma^\mu\Gamma^\nu\Gamma^i\chi_i\right)
\partial_\nu\phi\nn\\[5pt]
&&-\ft14 e^{-4\alpha\phi/3} F_{\mu\nu}^i \Big[
{\bar\psi}^\mu\Gamma_i \psi^\nu-
\pb_\rho(\Gamma^{\mu\nu\rho}\Gamma^i\Gamma^k
+\Gamma^{\mu\nu}\Gamma^\rho\delta_i^k)\chi_k\nn\\[5pt]
&&+\ft12
{\bar\chi}_j(\Gamma^{ijk}-4\Gamma^j\delta^{ki})
\Gamma^{\mu\nu}\chi_k\Big]
\nn\\[5pt]
&&-\ft1{32}g e^{4\alpha\phi/3}C_{i,\,jk}
\Big(\pb_\mu\Gamma^{\mu\nu}\Gamma^{ijk}\psi_\nu
-4\pb_\mu\Gamma^\mu\Gamma^{ij}
\chi^k-4{\bar\chi}^j\Gamma^{ki\ell}\chi_\ell
\nn\\[5pt]
&&+2{\bar\chi}^j\Gamma^i\chi^k+4{\bar\chi}^i\Gamma^j\chi^k\Big)
-\ft1{24}me^{2\alpha\phi/3}\left(\epsilon_{abc} X^{abc}
+\epsilon_{ijk}X^{ijk}\right)\nn\\[5pt]
&&+ \ft18
e^{2\alpha\phi/3}\left(\epsilon_{abc}
X^{abi}-\epsilon^{ijk}X_{cjk}\right)
B_i^c\ , \la{lag3} \eea
where $\psi_\mu= e_\mu^a\psi_a$, $\Gamma_\mu=e_\mu^a\Gamma_a$ and
$X^{ABC}$ are defined in \eq{x}. The explicit form of the last
four terms will be given shortly. It is convenient to leave them
in this form in comparing the B-field equation that follows from
\eq{pa} with ${\cal L}_F$ as given above, with the $S^3$
reduction of \eq{be}. The former is
\be \epsilon^{\mu\nu\rho}\Big({\cal D}_\mu B_\nu^\alpha L_\alpha^i
-mF_{\mu\nu}^i+\ft14
e^{4\alpha\phi/3}\epsilon_{\mu\nu\sigma}T^{ij}B^\sigma_j\Big)=\ft18
e^{2\alpha\phi/3}T^{ij}\left(\epsilon^{\mu\nu\rho} X_{\nu\rho
j}-\epsilon_{jk\ell} X^{\mu jk}\right)\ . \la{be2}\ee
Comparing this result with the $SU(2)$ reduction of the B-field
equation \eq{be}, we find a perfect agreement by choosing
\bea Y_\a^i=-\ft1{16}e^{-2\alpha\phi/3}
\epsilon_{abc}X^{bci}\ ,\nn\\[5pt]
X=\ft1{48} m^{-1}e^{-2\alpha\phi} \epsilon^{abc}X_{abc}\ .\la{sol}
\eea
The ansatz \eq{hcomp} then takes the form
\bea &&\hh^+_{abc} = m\, e^{\a\phi}\,
\ep_{abc}-\ft18 e^{-\alpha\phi}X^{abc} \,,\qquad \hh^+_{abi} = -
e^{-\fft13 \a\phi}\, \ep_{ab}{}^c \,
B_c^i-\ft18 e^{-\alpha\phi}X_{abi}\,,\nn\\[5pt]
&& \hh^+_{ijk} = m \, e^{\a\phi}\, \ep_{ijk}+\ft18 e^{-\alpha\phi}
X_{ijk}\,,\qquad \hh^+_{aij} = e^{-\fft13\a\phi}\, \ep_{ijk}\,
B_a^k+\ft18 e^{-\alpha\phi}X_{ija}\, .\label{hcomp2} \eea
The sign differences in the $X$-terms are essential for
the self duality of $H$. These terms are responsible
for the cancellation of the differentiated fermion terms that
arise on the right hand side of \eq{be}.

Having determined the quantities $X$, $Y_c^i$ and $Y_a^k$ in the
ansatz \eq{hcomp} for $\hh_{ABC}$, we can now express the
fermionic Lagrangian \eq{lag3} as follows:

\bea e^{-1}{\cal L}_F &=& -\ft12
{\bar\psi}_\mu\Gamma^{\mu\nu\rho}{\cal D}_\nu\psi_\rho
-{\bar\chi}_i\Gamma^i\Gamma^{\mu\nu}{\cal D}_\mu\psi_\nu
+\ft12 {\bar\chi}_i\Gamma^{ij}\Gamma^\mu{\cal D}_\mu\chi_j\nn\\[8pt]
&& +\ft12( {\bar\psi}_\nu\Gamma^\mu\Gamma^\nu\Gamma^i\chi^j
+{\bar\chi}_k\Gamma^{ki}\Gamma^\mu\chi^j) P_{\,\mu ij}
-\ft23\alpha\left({\bar\psi}_\nu\Gamma^\mu\Gamma^\nu\Gamma^i\chi_i\right)
\partial_\nu\phi\nn\\
&&+\ft12 e^{-4\alpha\phi/3} F_{\mu\nu}^i
\Big(\pb_\mu\Gamma^{\mu\nu\rho}\chi_i
+{\bar\chi}_k\Gamma^k\Gamma^{\mu\nu}\chi_i\Big)\la{fermionic}\\[5pt]
&&-\ft18 e^{-4\alpha\phi/3}G^+_{\mu\nu i}
\Big(2\pb^\mu\Gamma^i\psi^\nu-2\pb_\rho
\Gamma^{\rho\mu\nu}\Gamma^{ij}\chi_j
+{\bar\chi}_j\Gamma^{ijk}\Gamma^{\mu\nu}\chi_k\Big)\nn\\
&&-\ft1{32}g e^{4\alpha\phi/3}C_{i,\,jk}
\Big(\pb_\mu\Gamma^{\mu\nu}\Gamma^{ijk}\psi_\nu
-4\pb_\mu\Gamma^\mu\Gamma^{ij}\chi^k-4{\bar\chi}^j
\Gamma^{ki\ell}\chi_\ell\nn\\[5pt]
&&+2{\bar\chi}^j\Gamma^i\chi^k+4{\bar\chi}^i\Gamma^j\chi^k\Big)
+\ft12 me^{2\alpha\phi/3}\left(
\pb_\mu\Gamma^{\mu\nu}\sigma_1\psi_\nu
-{\bar\chi}_i\Gamma^{ij}\sigma_1\chi_j\right)\ ,\nn \eea
where
\be G^+_{\mu\nu i}= F_{\mu\nu i} + 2e
e^{2\alpha\phi}\epsilon_{\mu\nu\rho}B^\rho_i \ . \la{defg} \ee
Thus, the total $3D$ supergravity Lagrangian is given by \eq{pa},
\eq{bosonic} and \eq{fermionic}. As expected, the gravitino and
spinor kinetic terms are mixed. We have verified that they can be
simultaneously be diagonalized for any dimension $n$. In
particular, the field redefinitions which do the job for $n=3$ are
given by \footnote{The redefinition $\psi_\mu = \psi'_\mu -
\Gamma_\mu\Gamma^k\chi_k$ suffices to eliminate the mixing between
the fermion kinetic terms and puts the gravitino its canonical
form, while the kinetic term for $\chi_i$ remains non-diagonal in
the $SU(2)$ space. This kind of procedure was adopted in
\cite{ss8}.}
\bea
 \psi_\mu&=& \psi'_\mu -\ft12 \Gamma_\mu\Gamma^k\chi'_k\
,\nn\\[5pt]
\chi_i&=& -\ft12\Gamma^k\Gamma_i\chi'_k\ .\la{fr} \eea
The inverse transformation is
\bea
\psi'_\mu&=&\psi_\mu+\Gamma_\mu\Gamma^k\chi_k\ ,\nn\\
\chi'_i&=&\Gamma_{ij}\chi^j\ . \la{def12b} \eea
The kinetic terms become diagonal in terms of the primed fields
and in particular the spinor kinetic term becomes
$-\ft12{\bar\chi}'^i\Gamma^\mu{\cal D}_\mu\chi'_i$.


\subsection{The $3D$ Supersymmetry Transformations}


The reduction of the gravitino transformation rule has been carried
out already in \cite{lps}. Here, we shall also perform the $S^3$
reduction of the supersymmetry transformation rules of the graviton
and the self-dual field strength, thereby obtaining all the
transformation rules up to cubic fermion terms.

Let us begin with the combined supersymmetry and Lorentz
transformations of the vielbein $N=(1,0)$ supergravity in six
dimensions:
\be{\widehat e}_A{}^M\delta {\widehat e}_{MB}=
\widehat{\bar\epsilon}\, \Gamma_B \hp_A +{\widehat \lambda}_{AB}\
,\la{1}\ee
In order that the 3D gravitino transformation takes the form
$\delta \psi_a(x)=D_a \epsilon(x)+\cdots$, we need to make
the ansatz
\be {\widehat\epsilon}(x,y)=e^{\alpha\phi/2}\epsilon(x)\
.\la{2}\ee
The $ij$ projection of \eq{1}, and its trace, then yield
\bea L_i{}^\alpha\delta L_{\alpha j} &=&
{\bar\epsilon}\Gamma_{(i}\chi_{j)} -\ft13\delta_{ij}\left(
{\bar\epsilon}\Gamma^k\chi_k\right)+\lambda_{ij}\ ,
\la{3}\\[5pt]
\delta\phi &=& -\ft1{\alpha}\, {\bar\epsilon}\Gamma^i\chi_i\
,\la{4} \eea
where the composite $SO(3)$ transformation parameter
$\lambda_{ij}$ in $3D$ theory is defined as
\be \lambda_{ij}={\widehat\lambda}_{ij}-\eb\Gamma_{[i}\chi_{j]}\ .
\la{5} \ee
The $(ia)$ projection of \eq{1} gives
\be {\widehat \lambda}_{ia}={\bar\epsilon}\Gamma_a\chi_i\ ,
\la{6}\ee
which is the required Lorentz transformation to maintain the
triangular gauge implied by \eq{vielbein}. Next, the $(ai)$ projection
of \eq{1} gives
\be \left(\delta A_\mu^\alpha \right)L_\alpha^i = -\ft12 g
e^{4\alpha\phi/3}
\left({\bar\epsilon}\Gamma^i\psi_\mu+\eb\Gamma_\mu \chi^i\right)\
, \la{7} \ee
where we have used \eq{5}. Finally, the $ab$ projection of
\eq{1} yields
\be \delta e_\mu^a= {\bar\epsilon}\Gamma^a\psi_\mu
+\left({\bar\epsilon}\Gamma^i\chi_i\right) e_\mu^a
-\lambda^{ab}e_{\mu b}\ ,\la{8} \ee
where the second term disappears if we were to make the field
redefinition \eq{fr} which diagonalizes the fermionic kinetic
terms as discussed earlier, and the Lorentz transformation
parameter in $3D$ theory is $\lambda_{ab}\equiv
{\widehat\lambda}_{ab}$.

The reduction of the gravitino transformation rule
\be \delta \hp_A = {\widehat\nabla}_A\,{\widehat \epsilon}+\ft14
\hh_A^{+\ CD}\Gamma_{CD}{\widehat \epsilon}\la{9} \ee
is straightforward, given the expressions \eq{sc} for the spin
connection. The Lorentz transformations are not written down as
they do not give rise to induced supersymmetry transformations to
lowest order in fermions. The reduction of the above
transformation rule gives \cite{lps}
\bea \delta\psi_\mu &=& {\cal D}_\mu \epsilon +\ft12\alpha
\Gamma_\mu\Gamma^\nu\partial_\nu\phi -\ft12
e^{-4\alpha\phi/3}F_{\mu\nu}^i \Gamma^\nu\Gamma_i\epsilon\nn\\
&&+\ft12
e^{2\alpha\phi/3}B_\nu^i\Gamma^\nu\Gamma_\mu\Gamma_i\sigma_1\epsilon-\ft12
me^{2\alpha\phi}\Gamma_\mu\sigma_1\epsilon\ , \la{10}\\[5pt]
\delta \chi_i &=& \ft12 \left(P_{\mu
ij}-\ft13\alpha\delta_{ij}\partial_\mu\phi\right)
\Gamma^j\Gamma^\mu\epsilon +\ft14 g e^{4\alpha\phi/3}
\left(T_{ij}-\ft12\delta_{ij}T\right) \Gamma^j\sigma_1\epsilon
\nn\\
&& +\ft14 e^{-4\alpha\phi/3} F_{\mu\nu}^i \Gamma^{\mu\nu}\epsilon
+\ft12 e^{2\alpha\phi/3}
B_\mu^k\Gamma_k\Gamma_i\Gamma^\mu\sigma_1\epsilon\ . \la{chi}\eea
There remains the transformation rule for $B_\mu^\alpha$. To this
end, it is convenient to first define
\be h_{ABC} \equiv \hh_{ABC}^{+ cov} -\ft14 \epsilon_{ABCDEF}
\hp^D\Gamma^E\hp^F\ ,\la{hhp}\ee
where $\hh_{ABC}^{+ cov}$ is the supercovariant self-dual field
strength defined as
\be \hh_{ABC}^{+cov}=\hh_{ABC}^+
-\ft32\hp_D\{\Gamma^{DE},\Gamma^{ABC}\}\hp_E\ .\ee
It follows from \eq{hcomp2} that
\be h_{abi} = -e^{-\alpha\phi/3} \epsilon_{abc} B^c_i\ . \la{defb}\ee
The advantage of working with $h_{ABC}$ is that its transformation
rule is somewhat simpler to compute. Using \eq{dh2}, and recalling
that we work up to cubic fermions, we find that the combined
supersymmetry and Lorentz transformation of $h_{ABC}$ is given by
\bea \delta h_{ABC}&=&\ft32
{\widehat\epsilon}\,\Gamma_{[A}\hp_{BC]}{(\omega(e)}) -\ft32
{\widehat\epsilon}\Gamma^D\hp_{[A}\hh_{BC]D}^+
-\ft12\epsilon_{ABCDEF}\hp^D\Gamma^E\delta\hp^F\nn\\
&&+3{\widehat\lambda}_{[A}{}^D h_{BC]D}\ . \la{dhhp}\eea
From this equation, using \eq{defb} and the formulae provided above
for various quantities occurring in this equation, including the
compensating $SO(3)$ transformation \eq{5}, we find teh supersymmetry
transformation rule
\bea
\left(\delta B_\mu^\alpha\right) L_\a^i &=& e^{-2\alpha\phi/3}
\left( 2{\bar\chi}_j \Gamma^{ij}\delta\psi_\mu
+\epsilon^{ijk}{\bar\chi}_j \Gamma_\mu \delta\chi_k\right)\la{deltaB}\\[5pt]
&&-\ft12 m e^{4\alpha/3} {\bar\epsilon} \Gamma_\mu\chi^i
+\ft14 e^{\alpha\phi/3}\left( \epsilon_{\mu\nu\rho}{\bar\epsilon}
\Gamma^i\psi^{\nu\rho}-2{\bar\epsilon}\Gamma_{\mu\nu}\psi^{\nu i}\right)
\nn\\[5pt]
&& +{\bar\epsilon}\left( g_{\mu\nu}(-\Gamma^i\chi^j+\ft12
\delta^{ij}\Gamma^k\chi_k)
-\Gamma_{\mu\nu}\Gamma^{ijk}\chi_k+\Gamma_{(\mu}\psi_{\nu)}
\delta^{ij}+\ft12 \Gamma_{\mu\nu\rho}\Gamma^{ij}
\psi^\rho\right)B^\nu_j\nn \ . \eea

To summarize, the supersymmetry transformation rules of the $3D$
supergravity are given by \eq{3}, \eq{4},\eq{7},\eq{8},\eq{10},
\eq{chi} and \eq{deltaB}. It is straightforward, though
considerably tedious, to perform the redefinitions \eq{fr} if one
wishes to work with diagonalized fermionic kinetic terms. We also
note that the transformation rule for $B_\mu^\a$ is rather
complicated and it has an unusual form. This is due to the fact
that this field originates directly from the 3-form field
strength, instead of its potential. There are many ways to express
this transformation rule, and a better understanding of how to do
so in a manner that would be natural from a direct
three-dimensional construction would be desirable.


\subsection{Yang-Mills and Matter Vector Fields With Opposite Helicity}


In this section we take a closer look at the coupled field
equations for the $SU(2)$ Yang-Mills vector fields and the triplet
of matter vector fields. The bosonic part of their field equations
are given by
\bea &&{\cal D}^b(e^{-\fft83\a\phi}\, F_{ab}^i) =
-e^{-\fft83\a\phi}\, P^{b\,ij} F_{ab\,j}\, + \ft14 g^2\,
\ep^{ijk}\, T_k{}^{\ell}\, P_{\,a j\ell} - 4 m\, e^{\ft43\a\phi}\,
B_a^i + 2 \ep^{ijk}\, \ep_{abc}\, B^b_j\, B^c_k\,,\nn\\[5pt]
&& \epsilon^{abc} \left( D_b B_c^\a -mF_{bc}^\a \right)L_\a^i =
g\,e^{\ft43\a\, \phi}\, T^{ij}\, B^a_j\, .\label{abeqn} \eea
To analyze the representation content of the vector fields, it
suffices to examine their linearized equations of motion.  In
doing so, we set $g=4m$ which implies that the AdS$_3$ vacuum
arises with vanishing scalar fields. (Note that $R_{ab}=-2m^2
\eta_{ab}$, and hence, $m$ is the inverse AdS$_3$ radius). The
linearized equations of motion then then take the form
\be d{*F}^i = -4m\, F^i  + \ft12 G^i\,,\qquad B^i = \ft{1}{2m}\,
{*G^i} -{*F^i}\,,\label{ablinear} \ee
where $G^i=dB^i$.  We now define the one-forms $A_1^i$ and $A_2^i$
by
\be A^i = A_1^i + A_2^i\,,\qquad B^i =m\, (4A_1^i + A_2^i)\,. \ee
With these definitions, it follows from (\ref{ablinear}) that
\be d{*F_1^i}=4m\, F_1^i\,,\qquad d{*F_2^i}=-2m\, F_2^i\ , \ee
where $F_1^i=dA_1^i$ and $F_2^i=dA_2^i$. These equations can be
derived from an action which is the sum of two actions each one
containing a Yang-Mills kinetic term and Chern-Simons mass terms
with opposite signs:
\be e^{-1}{\cal L}=-\ft14 F_{1\,ab}F^{ab}_{1}+m
\e^{abc}F_{1\,ab}A_{1\,c}- \ft14 F_{2\,ab}F^{ab}_{2}-\ft12 m
\e^{abc}F_{2\,ab}A_{2\,c}\ .\la{l2} \ee
The sign of the topological mass term dictates the sign of the
spin-1 helicity, and thus we see that the vector fields $A_1$ and
$A_2$ describe spin-1 fields with opposite helicity \cite{dkss}.
This result is in exact agreement with that of \cite{dkss} (with
AdS$_3$ radius set to 1), where these spin-1 fields arise as a
subset of those which come from an $SO(4)$ gauge invariant KK
reduction from six dimensions on $S^3$. As explained in
\cite{dkss}, the Lagrangian \eq{l2} is only meant to be
interpreted at the free level (which is sufficient for deducing
the representation content) and that the interacting Lagrangian
cannot be written in this way. Indeed, that would have implied an
$SO(4)$ gauge symmetry, while we know that the full theory is only
$SU(2)$ gauge invariant.


\section{Conclusions}


     In this paper, we have completed the construction of the
three-dimensional supergravity obtained by an $S^3$ group-manifold
reduction of pure $(1,0)$ chiral supergravity in six dimensions.  The
bosonic action, and the fermionic transformation rules, had been
obtained previously in \cite{lps}.  The resulting three-dimensional
supergravity has $SU(2)$ Yang-Mills fields with topological mass
terms, and a triplet of massive vector fields in the adjoint
representation of $SU(2)$, together with six scalars described by the
coset $GL(3,R)/SO(3)$.

    In our reduction we neglected the quartic fermion terms in the
six-dimensional Lagrangian, and the associated cubic-fermion terms in
the supersymmetry transformation rules.  However, we did keep all
contributions following from the quadratic fermion terms in the
Lagrangian, meaning in particular that we kept all the bilinear
fermion contributions in the bosonic equations of motion.  It was
therefore necessary to work with the full self-duality condition for
the six-dimensional 3-form field, including the fermion bilinear
terms. Thus although the procedure for performing a group-manifold
reduction is a mechanical one, there were considerable subtleties in
this case arising from the reduction of the full self-duality
condition.

    Although the six-dimensional theory cannot have a Lagrangian
formulation, owing to the self-duality of the 3-form, the reduced
three-dimensional theory {\it does} have a Lagrangian formulation.
Furthermore, it is unnecessary to introduce potentials for the
three $B$-type matter vector fields defined in (\ref{hcomp2})
which arose directly from the self-dual 3-form field strength (as
opposed to its potential) in six dimensions.  This is because the
six-dimensional self-duality is transformed into an
``odd-dimensional self-duality'' in $D=3$, which admits a
Lagrangian formulation even though the equations of motion are of
first order.

   The matter-coupled AdS$_3$ supergravity that we have obtained here
has a specific field content and rather elaborate interactions which
would be difficult to guess from a direct construction in three
dimensions. It is of interest, however, to find a systematic way of
directly constructing not only this model but its generalizations with
arbitrary field content, as well as higher supersymmetries.
Significant progress in this direction has bene made in \cite{nsr}.
In particular, the model we have constructed has been suggested to
correspond to a Chern-Simons type of supergravity with CS gauge group
a semidirect product of $SO(3)$ with a 6-parameter group generated by
3 abelian and 3 nilpotent generators.  This is coupled to 12 scalars
that parametrize $SO(4,3)/SO(4)\times SO(3)$.  In \cite{nsr}, a
procedure is described in which 3 of the scalars are dualised to what
we call $B$-type vector fields, and 3 scalars are gauged away, thereby
leaving 6 scalars that parametrize $GL(3,R)/SO(3)$. So far, the
bosonic action has been provided in \cite{nsr}, and while it seems to
have a structure similar to our bosonic action, the exact equivalence
remains to be shown.

     In this paper, we have focussed on the $S^3$ group-manifold
reduction of pure $(1,0)$ six-dimensional supergravity, which is, of
course, anomalous.  Since one of the principal motivations for
studying such reductions is to obtain lower-dimensional AdS
supergravities that are embedded in string theory, it is natural to
extend our work by taking an anomaly-free six-dimensional theory as
the starting point for the $S^3$ reduction.  There exists a variety of
such six-dimensional anomaly-free theories, including $(1,0)$ examples
with appropriate matter multiplets, which can be obtained from K3
reductions of the heterotic string, and the $(2,0)$ theory coupled to
21 tensor multiplets, which comes from the type IIB string reduced on
K3.  All the resulting three-dimensional supergravities will
presumably be encompassed within the general class that has been
proposed in \cite{nsr}.  While it is useful to have a direct
construction of the most general matter-coupled AdS supergravities in
three dimensions, it is also of importance to establish a connection
with those which follow from $S^3$ reduction, in order to see how they
are embedded in string theory.  This would be of particular
importance, for example, for studying the AdS$_3$/CFT$_2$
correspondence.

\newpage


\begin{appendix}


\section{Conventions}


We use the signatures $\eta_{AB}=(-+\cdots +)$ and $
\eta_{ab}=(-++)$, and use the $\Gamma$-matrices
\bea \Gamma^a &=& \gamma^a\times 1 \times \sigma^1\ ,\quad\quad
\{\gamma_a,\gamma_b\}=2\eta_{ab}\ , \quad\quad
\gamma^{abc}=\epsilon^{abc}\ ,\\
\Gamma^i &=& 1\times \gamma^i\times\sigma^2\ ,\quad\quad
\{\gamma_i,\gamma_j\}=2\delta_{ij}\ ,\quad\quad
\gamma^{ijk}=-i\epsilon^{ijk}\ ,\nn\\
\Gamma_7 &=& 1\times 1\times\sigma^3\ ,\quad\quad\
\Gamma_{A_1A_2...A_6}=\epsilon_{A_1A_2...A_6}\Gamma_7\ ,\nn\\
C&=& (i\sigma^2)\times(i\sigma^2)\times \sigma^1\ . \eea
The 6D spinors are symplectic Majorana-Weyl and $\Gamma_7
\epsilon=\epsilon$. The $Sp(1)$ doublet index $r=1,2$ is
suppressed. In all the fermionic bilinears the north-west
south-east contraction rule is understood, and
$\epsilon_{rs}\epsilon^{rt}=\delta_s^t$. With the $Sp(1)$ indices
suppressed, we have the symmetry property
\be {\bar \psi}_1 \Gamma^{A_1\cdots A_n}\psi_2= (-1)^n {\bar
\psi}_2 \Gamma^{A_n\cdots A_1}\psi_1\ , \ee
for any two anticommuting symplectic Majorana spinors $\psi_1$ and
$\psi_2$.


\section{Identities for Curvature with Torsion}


We begin by noting the usual vielbein postulate

\be
\partial_\mu e_\nu^a+\omega_\mu{}^{ab}(e)\, e_{\nu b}
-{\rho\brace \mu\nu}\, e_\rho^a =0\ , \la{vp1}\ee
from which one solves for the spin connection
\be \omega_{\mu ab}(e) = \Big(e^\nu_ a
\partial_{[\mu}e_{\nu]b}+\ft12\, e^\rho_a e^\nu_b e_\mu^c
\partial_\nu e_{\rho c} -  a \leftrightarrow b
\Big) \ee
We also require the condition ${\cal D}_\mu(\Gamma,\omega_+)
e_\nu^a=0$, where $\omega_{\pm\,\mu}{}^{ab}$ is given by \eq{opm}.
This condition explicitly takes the form
\be \partial_\mu e_\nu^a+\omega_\mu^{+\ ab} e_{\nu b}
-\Gamma^\lambda{}_{\mu\nu}\, e_\lambda^a = 0.\la{vp2} \ee
Taking the antisymmetric part and using \eq{vp1} we learn that
\bea \Gamma^\lambda{}_{[\nu\mu]} &=& S_{\mu\nu}{}^\lambda
+\hh_{\mu\nu}^{+\ \lambda}\ ,\\[5pt]
S_{\mu\nu}{}^\lambda &=& -\ft12\,\pb_\mu\Gamma^\lambda\psi_\nu\ .
\eea
The antisymmetric part of $\Gamma^\lambda{}_{\mu\nu}$ has the
interpretation of torsion. In addition to the usual fermionic
bilinear contribution to torsion, here the supercovariant
self-dual field strength $\hh_{\mu\nu\rho}^+$ also appears as a
contribution to torsion.

Subtracting \eq{vp1} from \eq{vp2}, we also learn that
\bea \Gamma^\lambda{}_{\mu\nu} &=& {\lambda\brace \mu\nu}-
K_{\mu\nu}{}^\lambda\ ,\la{gc}\\[5pt] \
 K_{\mu\nu}{}^\lambda &=& \kappa_{\mu\nu}{}^\lambda
+\hh_{\mu\nu}^{+\ \lambda}\ . \eea
Note that $K_{\lambda(\mu,~\nu)}=0$, where
$K_{\mu\nu,\,\rho}\equiv K_{\mu\nu}{}^\lambda g_{\lambda\rho}$.
Observe also that we can write \eq{gc} as
\bea \Gamma^\lambda{}_{\mu\nu} &=&
{\widehat\Gamma}^\lambda{}_{\mu\nu}- \hh_{\mu\nu}{}^\lambda\
,\\[5pt]
{\widehat\Gamma}^\lambda{}_{\mu\nu} &\equiv& {\lambda\brace
\mu\nu}-\kappa_{\mu\nu}{}^\lambda\ . \eea
Using these equations, and recalling the definition of
$\omega_{\pm\,\mu}{}^{ab}$ given in \eq{opm}, we find that the
vielbein postulate \eq{vp2} can also be written as
\be \partial_\mu e_\nu^a+\ho_\mu{}^{ab} e_{\nu b}
-{\widehat\Gamma}^\lambda{}_{\mu\nu}\, e_\lambda^a = 0.\la{vp3}
\ee

Next, we define the Riemann tensors
\bea
R^\sigma{}_{\rho,\,\mu\nu}(\Gamma) &=& \left(
\partial_\mu\Gamma^{\sigma}{}_{\nu\rho}
+\Gamma^{\sigma}{}_{\mu\lambda}\Gamma^{\lambda}{}_{\nu\rho} - \mu
\leftrightarrow \nu \right)\ ,\\[5pt]
R_{\mu\nu}{}^{ab}(\omega) &=& \left(\partial_\mu\omega_\nu{}^{a
b}+\omega_\mu{}^{ac}\omega_{\nu c}{}^b- \mu \leftrightarrow \nu
\right)\ . \eea

Using the vierbein postulates \eq{vp2} and \eq{vp3}, we find that
\bea R^\sigma{}_{\rho,\,\mu\nu}(\Gamma) &=&
R_{\mu\nu}{}^{ab}(\omega^+)\, e^\sigma_a e_{\rho b}\ ,\la{rr1}\\[8pt]
R^\sigma{}_{\rho,\,\mu\nu}({\widehat\Gamma}) &=&
R_{\mu\nu}{}^{ab}(\ho)\, e^\sigma_a e_{\rho b}\ .\la{rr2} \eea
Next, we define $R_{\rho\sigma,\,\mu\nu}
(\Gamma)=g_{\rho\lambda}R^\lambda{}_{\sigma,\,\mu\nu}(\Gamma)$ and
recalling \eq{gc} we find:
\bea
R_{\rho\sigma,~\mu\nu}(\Gamma) &=&
R_{\rho\sigma,~\mu\nu}(\{\})+
2K_{[\mu\nu]}{}^\lambda K_{\lambda\sigma,~\rho}\nn\\
&& +\Big( K_{\mu\sigma}{}^\lambda K_{\nu\lambda,~\rho}+
D_\mu(\Gamma)K_{\nu\rho,~\sigma}-\mu\leftrightarrow \nu\Big)\
.\la{gr1} \eea

It follows that $R_{(\rho\sigma),~\mu\nu}(\Gamma)=0$. The
following form of the above equation is also useful:
\bea R_{\rho\sigma,~\mu\nu}(\Gamma) &=&
R_{\rho\sigma,~\mu\nu}({\widehat \Gamma})+
2D_{[\mu}(\Gamma)\hh^+_{\nu]\rho\sigma}-
\hh^+_{\mu\nu}{}^\lambda\hh^+_{\rho\sigma\lambda}\ .\la{gr2} \eea
In obtaining this result, we have used the identity
\be \hh^+_{[\mu\nu}{}^\lambda \hh^+_{\rho]\sigma\lambda}=0\ . \ee
 Next, we define the Ricci tensor as
\be R_{\mu\nu} (\Gamma)\equiv g^{\rho\sigma}
R_{\mu\rho,~\nu\sigma}(\Gamma)\ .\ee
By direct computation we find from \eq{gr1} that
\be R_{[\mu\nu]}(\Gamma)={\cal D}_\lambda(\Gamma)\hh_{\mu\nu}^{+\
\lambda} +\ft12 \pb_a\Gamma^a\hp_{\mu\nu} +2\pb^a\Gamma^b
\psi_{[\mu} \hh^+_{\nu]ab}\ . \la{ricci}\ee

\end{appendix}

\newpage


\end{document}